\documentstyle[psfig,sprocl]{article}
\def\be{\begin{equation}}
\def\ee{\end{equation}}
\def\bea{\begin{eqnarray}}
\def\eea{\end{eqnarray}}

\begin{document}

\title{Deuterium and Helium Absorption at High Redshift:
Mapping the Abundance, Density and Ionization of Primordial Gas}

\author{Craig J. Hogan}

\address{Astronomy and Physics Departments,
University of Washington\\PO Box 351580, Seattle, WA 98195, USA}
  

\maketitle\abstracts{
Spectra of quasars at high redshift
with high resolution and  high signal-to-noise   allow 
in favorable circumstances detection of absorption by
 deuterium  and  ultimately measurement of its primordial
abundance. Ultraviolet spectra of high redshift quasars allow
measurement of absorption by the most abundant cosmic
absorber, singly ionized helium, thereby mapping 
 gas even in the most rarefied cosmic
voids.  These new techniques already provide significant 
constraints on cosmological models but will soon become
much more precise. }

\section{Absorption by Primordial Gas}

Soon after quasars were discovered in the 1960's it was
realized (by Bahcall and Salpeter, Gunn and Peterson, Lynds
and others) that their exceptional brightness and simple, smooth
intrinsic spectra made their absorption spectra
 ideal probes of distant material
along the line of sight--- a unique opportunity to study
material at great distances and early times
in detail.  By 1980 landmark papers by Sargent
and collaborators had laid out a considerable detailed 
statistical knowledge of the absorption, with many important
implications for cosmology. The statistical properties of 
the absorption firmly established that it is caused by
cosmologically distributed foreground gas, mostly
by  clouds
constituting an intergalactic or protogalactic population.
Some fraction of the clouds with high column density and metal 
enrichment were identified with galaxies already in the process
of chemical evolution. During the 1980's,
catalogs of the highest column density ``damped''
absorbers (especially by  Wolfe and 
collaborators)  seemed to isolate a population that could
be readily identified as progenitors of modern galaxies.

Theoretical ideas for the absorbing clouds initially
explored a
very large space of physically  plausible populations
embedded within the wide variety of extant galaxy formation scenarios.
As the Cold Dark Matter paradigm for structure
formation  sharpened during the 1980's, Rees and others showed that
within this picture the absorbing clouds are a natural
accompaniment to   galaxy formation;
 they are the condensations of the primordial 
baryonic material during its early stages of collapse into 
the dark matter potentials.  It became clear that the 
study of the distribution and the enrichment of the gas
through absorption would provide one  of the richest 
and most precise  ways of viewing directly the process
of galaxy and structure formation and chemical evolution.

This promise has been realized 
in the last few years as the pace of progress has advanced
quickly both in observations and interpretation.  The Keck
telescope now provides a qualitatively new type of data: 
spectra from $z\approx 3$ with high resolution (that
is, better than the thermal widths of lines), with 
signal-to-noise of the order of 100. The Hubble
and HUT telescopes
provide  another qualitatively new type of data: 
spectra from $z\approx 3$ down to below the ${\rm He^+}$ Lyman-$\alpha$
line 304\AA\  in
the rest frame, at lower resolution but still comparable
to the earlier work on HI. These two developments now
allow (among many other things)
 the study of absorption not only by hydrogen and metal ions
but also  by the other important 
primordial elements in this high redshift gas, deuterium
and helium.

At the same time,
ideas about the cosmic gas distribution have sharpened 
quantitatively in recent years, due to hydrodynamic
simulations  which accurately 
predict the distribution and motion of matter in  
hierarchical models of galaxy formation
(Cen et al. 1994, Hernquist et al. 1996, Miralda-Escud\'e et al.  1996,
 Rauch et al. 1997,
Croft et al. 1997, Zhang et al. 1997; see also Bi and 
Davidsen 1997).  Departing from earlier analytic
models based on isolated clouds with symmetric geometries such
as spheres and slabs,   simulations of 
gravitational collapse   from nearly uniform gas (with linear gaussian noise)
into nonlinear structures
produce   dynamical  systems
with a complex geometry (poetically described as voids, pancakes,
filaments, knots) and no sharp 
 distinction between diffuse gas and clouds; similarly,
 the  complex  simulated  absorption spectra
reveal no sharp   distinction between lines and continuum. 
It is now possible to use absorption spectra to apply
statistical tests to CDM models which are in some 
ways cleaner than the traditional ones based on galaxies---
since the gas distribution and ionization
 is computed accurately up to the point where optical 
depths become large, and do  not depend much on uncertainties
from star formation and extinction. The cleanest predictions
concern the gas in the least dense regions, especially
in the voids.

In this context the accessibility of deuterium and helium absorption
opens up new types of cosmological tests.
The  primordial abundance of deuterium is 
a critical observational test of  cosmological theory, both
as a test of the basic Big Bang picture and
as a measure of the cosmic baryon density, a central
parameter of structure-formation models
(Peebles 1966, Walker et al. 1991, Smith et al. 1993,
Copi et al. 1995, Sarkar 1996, Hogan 1997).
Even though deuterium is detected in the Galaxy,
 high redshift
absorption probes its primordial abundance with better
control over the effects of chemical evolution,
and   provides an opportunity
to map the abundance in space and its evolution in time
in different environments.

Helium absorption is also interesting for its
primordial abundance, but more so for the insights
it provides about the density, distribution and ionization
of gas at high redshift.
Simulated spectra reveal that
gas  in the most underdense regions,
filling the bulk of the spatial volume,
 is so highly ionized that it
produces  absorption features  with  very low HI Lyman-$\alpha$ optical depth. 
The more  abundant absorbing ion ${\rm He^+}$  however produces
optical depths of the order of unity even in these regions, so its absorption 
is easily detectable, mapping  the distribution of
cosmic baryons  at the lowest densities.

\section{ Deuterium Abundance}
 
There are now about   eight plausible detections of 
extragalactic deuterium in the literature, reviewed recently for 
example in Hogan (1997). There is currently no case   where
the candidate deuterium feature can be identified positively
as such: in every instance, the data could be interpreted
as an HI cloud accompanied by another cloud blueshifted
by 82 km/sec with a much smaller column density.
 In the cases where we had thought this was impossible
 based on the narrow widths of candidate D absorption 
(Rugers and Hogan 1996ab), new and better data
now show that the features are not so narrow after all
(Tytler, Burles and Kirkman 1996, and Cowie et al,
private communication.)  Indeed the new data shows that
there must be at least some contamination by HI at the
DI Lyman-$\alpha$ feature because the velocity centroid is slightly 
displaced from the bulk of the hydrogen as measured
from the higher Lyman series lines.
Thus, the evidence for a high deuterium abundance
is not conclusive, but  is based on the anecdotal 
accumulation of several high estimates, some of which are 
 reported in the literature. I describe here briefly
some recent work (Rugers and Hogan 1997) to seek
a more reliable statistical estimate.
\begin{figure}[htbp]
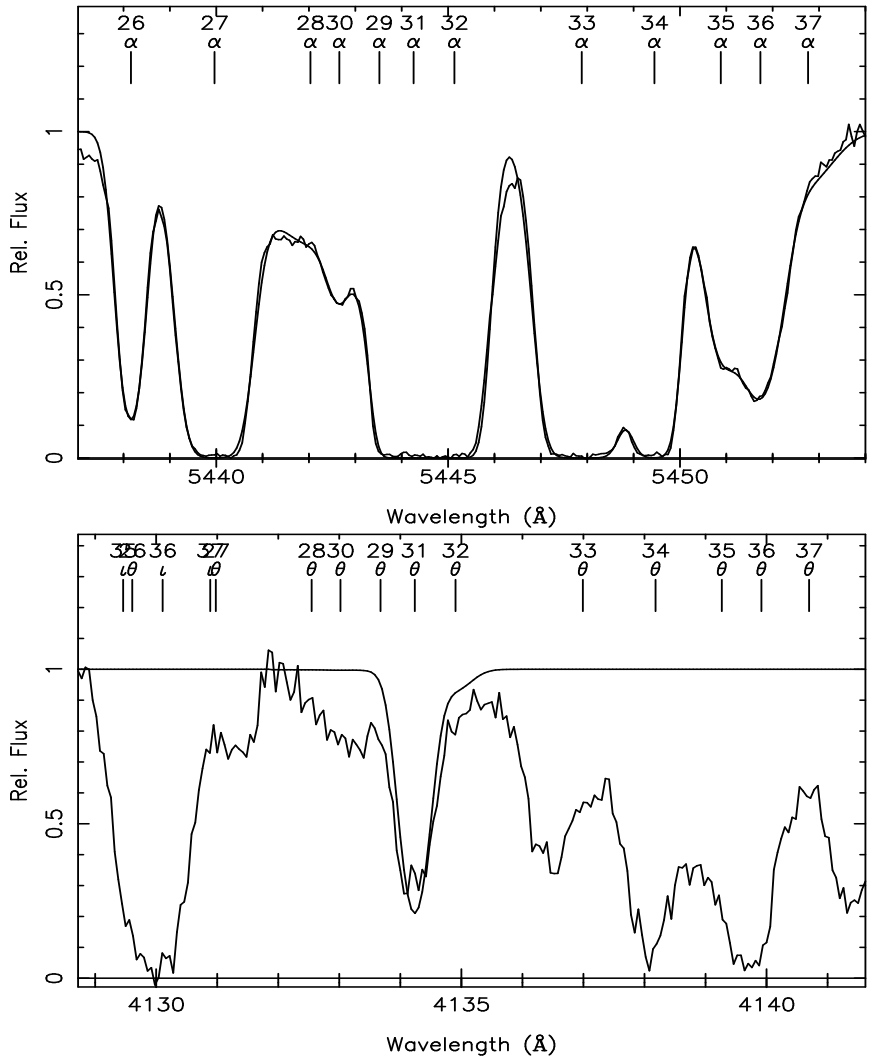

\psfig{figure=fig10a.ps,height=2.75in,width=4.5in,angle=270}
\psfig{figure=fig10e.ps,height=2.75in,width=4.5in,angle=270}
\caption{Lyman-$\alpha$ fit (top) for one of seven  new deuterium 
candidates, at $z=3.478401$ in  Q1422+230. The main 
HI absorption, component 31,
has redshift determined by the optically thin
Lyman-$\theta$ fit (below); the DI feature, component 30,
matches this redshift to within the fitting error
(difference $-3\pm 14$\ km/sec).
Hydrogen contamination (fitted mostly with component 28) is
superimposed on the deuterium feature, making
the abundance fit very uncertain,
$\log (D/H)=-3.53\pm 0.45$. It can nevertheless be used in
a statistical study.}
 \end{figure}
  
\subsection{Interloper statistics from  ``Pretend''
deuterium absorbers}
In principle, because of the monotonic destruction of deuterium
by chemical evolution, one should pay the most attention
to the highest measured abundances. But contamination by hydrogen lines
are bound to yield some spurious detections. Indeed, at some level 
there is always  some contamination, since there
is some hydrogen absorption at all redshifts--- the
problem is to quantify its effect on abundance estimates.

To estimate the contamination  
by hydrogen  interlopers, we explore a
statistical technique.  We have assembled a
 control sample  of ``pretend'' deuterium candidates
associated with hydrogen lines.
These candidates are
selected the same way the deuterium 
candidates are  selected, in
all respects but one: the  control sample candidates are
drawn from the red side of hydrogen absorbers, rather
than the blue side where the real deuterium feature appears.
To make a larger sample, the velocity bin 
to accept a pretend candidate is also larger 
than that for a deuterium candidate
(i.e., $[-60, -100]$\ km/sec rather than $[-82\pm 1\sigma]$\ km/sec).
The pretend candidates are all interlopers,
 drawn from  a
  population with the same statistical properties as the
 deuterium interlopers, including their joint correlations
in velocity, column density and width. 
 The properties of two samples can then be compared 
statistically,   using the Doppler
parameters and column densities from the line fits.

\begin{figure}[htbp]
\psfig{figure=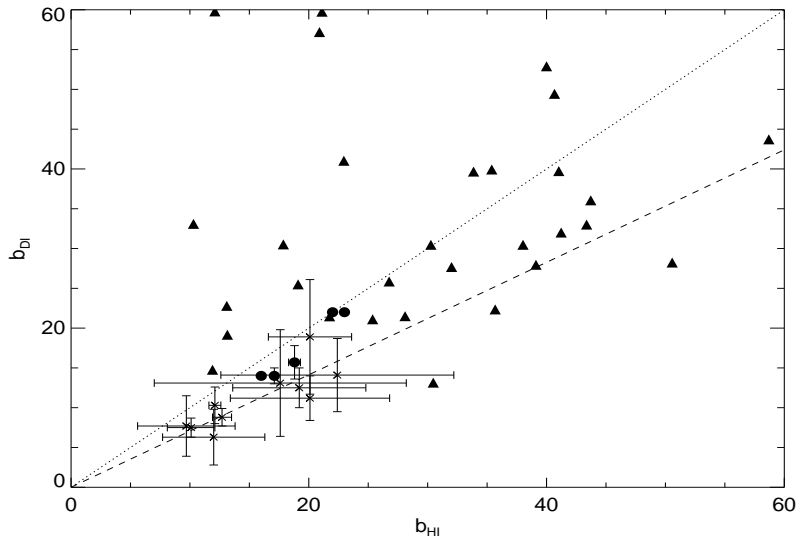,height=3in,width=4.5in,angle=90}
\caption{Fitted deuterium and hydrogen 
Doppler parameters   of real deuterium candidates 
(a uniform sample represented by crosses and 
published values by filled circles), 
compared with
 those in the sample  of pretend deuterium candidates, plotted
 as solid
triangles; errors in the latter are omitted
to enhance the clarity of the plot.
The two lines 
represent 
 the boundaries of the interval allowed for real deuterium,
between the extremes of thermal and turbulent broadening.
Unlike the pretend sample, all of the real deuterium candidates
are consistent with the allowed ratio. For the new 
uniform sample, no cut was applied based on Doppler parameter.}
 \end{figure}

For a fair comparison we have also assembled a uniform
 sample of 
deuterium absorbers from the same spectra as the 
pretend deuterium absorbers, based 
purely on redshift selection relative to HI and metal features.
 There are seven new candidates
with detectable deuterium features at the right redshift,
in addition to the previously published candidates
from Rugers and Hogan (1996ab).
The reason that they have not been published is that
they are not very convincing, or do not yield
a precise abundance. One new example
is shown in figure 1.

The fitted Doppler parameters (figure 2) show that the claimed 
deuterium detections are at least mostly real, even 
if individual cases are suspect. This is true both
for the new candidates and for the previously 
published ones. 
   The $D$ features have statistically narrower
 profiles than the pretend candidates,
 consistent with the deuterium
identification.

Since we believe the deuterium is statistically real,
we can derive a statistical
abundance. For the new sample of absorbers,
(that is, the uniform sample excluding previously
published values),
we obtain $\langle \log (D/H)\rangle= -3.75\pm 0.51$.
The reduced $\chi^2$ is 0.70, indicating 
consistency with a universal abundance.
This is certainly not the case for the sample
of pretend absorbers; even confining ourselves to
those with high HI column (so as to be more
directly comparable to the real ones), the
abundance of ``pretendium'' is 
$\langle \log (P/H)\rangle= -3.2\pm 2.6$,
with a reduced $\chi^2$ of 8.7.
Other statistics also display   differences.
 While the
real deuterium candidates display consistency
with a universal abundance (linear 
regression coefficient
between $N(D)$ and $N(H)$ of 1.00),
the pretend sample is a scatter plot
(linear 
regression coefficient 
between $N(P)$ and $N(H)$ of $0.55$.)
These statistics are reflected in the scatter plot shown 
in figure 3.
And the distributions of $N(D)$ and $N(P)$
also differ, in the sense that $P$ is 
not as common as $D$--- another way of 
saying that for the most part, interlopers
may be there but  
statistically  have a lower column density
than the  real
deuterium.



 \subsection{Current situation and future prospects}

Although deuterium is detected, its primordial
abundance is still uncertain by an order of magnitude.
At present, we have a very firm lower limit
$D/H\ge 2\times 10^{-5}$ on
primordial deuterium, from Galactic measurements
(Linsky et al. 1993, 1996),
as well as from some quasar absorbers 
(e.g. Tytler et al. 1996).
There is some evidence for a somewhat higher
lower limit ($D/H\ge 4\times 10^{-5}$) from quasar absorbers
(Songaila et al. 1996).  A higher abundance
than this is not clearly required by the data,
although there is some statistical evidence for it
and there is no very strong evidence against it,
since the low abundances are still found in just
a few cases where deuterium may have been destroyed.

Better even 
than a much larger statistical study would be to find a system where
a clear signature of deuterium can be proven,
or where the interloper probability is small.
A very promising possibility is a damped absorber  
in Q2206-199, at $z=2.559$, with a low metallicity
and very narrow lines (Pettini et al. 1994).
The interloper problem is much smaller here, both
because of the very high column and the low redshift.
The low redshift requires HST, but with STIS 
the entire Lyman series can be seen at once so
this is
a practical program, currently approved for
cycle 7.

\begin{figure}[htbp]
\psfig{figure=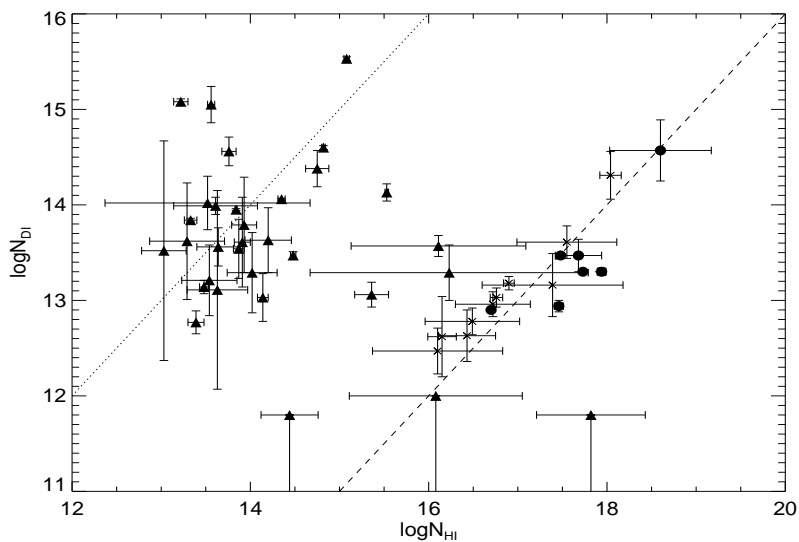,height=3in,width=4.5in,angle=90}
\caption{Column densities for fitted hydrogen components
and their deuterium counterparts
(crosses for the uniform sample, dots for other published
values), 
together with the same quantities for the control
sample of pretend candidates (triangles). The distributions
are clearly different, as confirmed by statistical tests.
Lines correspond to constant $D/H$.}

\end{figure}

\section{Resolving the Helium Lyman-$\alpha$ Forest:  Mapping
Intergalactic
Gas and Ionizing Radiation at $z\approx 3$}

There is certainly HI absorption between the 
identified lines of the Lyman-$\alpha$ forest.
As spectra of higher signal-to-noise ratio 
are obtained they reveal absorption of 
progressively lower optical depth.
In HI however, even at the highest S/N so far
available (about 100 at high resolution), 
the HI absorption does not yet fill redshift space.
Limits on HI continuous optical depth 
(or ``Gunn-Peterson effect'') provide useful constraints
on diffuse gas density, subject to uncertain
quantities such as the ionizing radiation field
(Giallongo et al. 1992,1994,1996).

Because of its higher ionization potential,
the  most  abundant absorbing
 ion in the universe is not HI but ${\rm He^+}$.
Even in  hard radiation fields near quasars
it is more abundant by a factor
$\eta$ between 10 and 20. In intergalactic space
the ratio is probably greater than 100, so
 in the
emptiest void  regions,  ${\rm He^+}$    produces much more easily detectable
optical depths even at modest signal-to-noise. It is therefore the 
best tool for mapping  the distribution of
cosmic baryons  at the lowest densities.
Absorption by ${\rm He^+}$ is also
the most direct probe of the hard ultraviolet 
cosmic radiation field, which can be predicted from
semiempirical models based on observed quasar
and absorber  populations
(Haardt and Madau 1996).  The spectral shape also influences
other observables such as the ratio of CIV to SiIV
(Songaila \& Cowie 1996), so information
from helium absorption allows information about
relative C and Si abundances to be derived.  In situations
where the ionizing spectrum is known, such as the near
proximity of a quasar,  ${\rm He^+}$ absorption can be compared
to HI absorption to extract
independent information about the primordial abundance of helium,
an important test of Big Bang Nucleosynthesis.

The difficulty of course is that the
Lyman-$\alpha$ transition of ${\rm He^+}$ is at 304\AA,
requiring observations both from space and at high redshift.
The first  detection of cosmic ${\rm He^+}$ absorption was made  in Q0302 by
Jakobsen et al. (1994)
using the Hubble Space Telescope Faint Object Camera (FOC).  They found an
absorption
edge and a large   continuous optical depth
at low resolution (10\AA), $\tau>1.7$, caused by
a mixture of lines and truly diffuse 
  $He^+$ Lyman-$\alpha$  absorption (Songaila et al. 1995).    A similar
observation has also been
made of the $z=3.185$ quasar PKS 1935-692, with a similar result
(Tytler \& Jakobsen 1996).

A significant improvement came from
  the Hopkins Ultraviolet Telescope (Davidsen et al. 1996),
which can reach shorter wavelengths and hence
lower redshift than HST, and also provides better 
resolution and wavelength calibration than FOC.
Davidsen et al. observed     the $z=2.72$
quasar  HS 1700+64 and found an $He^+$ edge   close enough
to the predicted redshift to rule out the possibility
of foreground HI as an important contaminant.
They also found that the flux below the edge is not
consistent with zero, and measured accurately a
mean optical depth, $\tau=1.00\pm 0.07$.
The smaller absorption at lower redshift  reflects   the increasing ionization
of ${\rm He^+}$  
 with time, the thinning due to the expansion,  and   the conversion
of diffuse gas into clouds.

New observations of Q0302 
(Hogan et al. 1997) were made
to improve both the wavelength calibration and
resolution of the ${\rm He^+}$ absorption,
with enough sensitivity to correlate usefully 
with the HI absorption (see figure 4). 
We can now
explore in detail the relative contributions of clouds
and diffuse gas, as well as measuring  independently the
ionizing radiation field  and the helium abundance.
We   find significant ${\rm He^+}$ absorption from  HI clouds
(with optical depth of the order of unity)
but also comparable ${\rm He^+}$ absorption
 even in redshift intervals where the best Keck spectrum
reveals no detectable HI; thus we   directly  measure
absorption attributable separately to both the clouds
and the diffuse gas.
Our spectrum  suggests
nonzero flux at all wavelengths, which 
constrains the ionizing background spectrum and  leads
to an upper limit on the density of diffuse gas.
Absorption 
from gas near the quasar,  where the incident spectrum is 
known approximately from the direct measurement of the quasar spectrum,
allows independent constraints on the density and helium abundance
of the gas.

The main new conclusions from the current data are:
1. The ${\rm He^+}$ Lyman-$\alpha$ forest is detected,
and indeed discrete clouds identified in HI are responsible for the main
${\rm He^+}$ absorption edge
in the spectrum;
2. The ``diffuse'' (redshift-space-filling)  
medium is also detected, and must have  a low density ($\Omega\le 0.01
(h/0.7)^{-3/2}$) consistent with standard primordial
nucleosynthesis and models of early gas collapse into protogalaxies;
3. The intergalactic ionizing spectrum is soft ($\eta\ge 100$),
although the intergalactic helium is probably mostly doubly ionized
by $z=3.3$; 4. The helium abundance is within a factor of
a few of standard Big Bang predictions, over a large volume
of space at high redshift. We expect that these conclusions will
be made more general and precise with the new 2-dimensional
spectroscopic capability of HST/STIS.
\section*{Acknowledgments}
Keck spectra used in this work were generously shared by 
L. Cowie and collaborators.
I am grateful for many useful discussions with participants in
the 1996 workshop on Nucleosynthesis at the Institute for Nuclear
Theory in Seattle, funded by DOE.
This work was supported   by NASA and the NSF
at the University of Washington.

 \begin{figure}[htbp]
\psfig{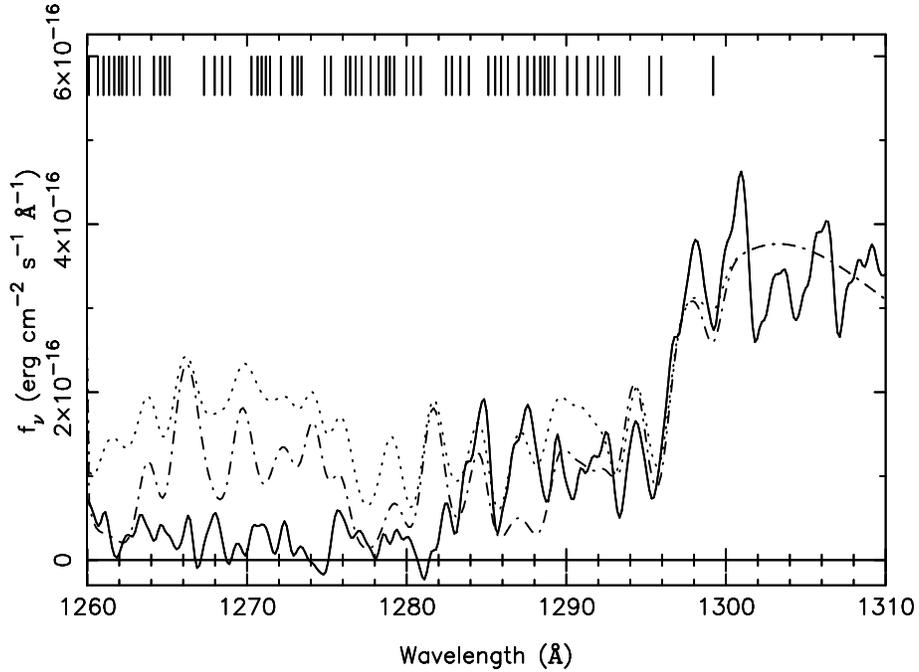}
\caption{Detail of HST spectrum of 0302-003 near the 
${\rm He^+}$ edge.
The HST spectrum  is overlaid with a model spectrum
predicted on the basis of the
model  distribution of HI derived from
a Keck spectrum of the HI Lyman-$\alpha$ forest.
   Ticks indicate the
fitted  HI velocity components from the Keck spectrum. Doppler
parameters and column densities from the fit were
used to predict the  ${\rm He^+}$ absorption spectrum at the GHRS resolution.
Two predictions are shown: 
dotted and dot-dash curves corresponding
to   ${\rm He^+/HI}$  ratios $\eta =20$ and 100 respectively,
both models assuming pure turbulent broadening, $b_{He^+}=b_{HI}$.
The  HST and Keck
spectra appear to show corresponding
absorption features, including  the
main ${\rm He^+}$ edge.
Near the quasar, a hard spectrum
with  $\eta=20$ is probably sufficient to explain
the absorption features  entirely with clouds. A large
 $z$-filling
optical depth is only allowed outside the 
proximity of the quasar, 
presumably because the diffuse gas nearer the
quasar is doubly ionized.
There is significant
${\rm He^+}$ opacity ($\tau_{GP} > 1.3$)
 even at the redshift of the conspicuous HI
Lyman-$\alpha$  forest void near 1266\AA,
indicating the presence of diffuse gas between identified
HI absorbers. At the same time there is significant nonzero flux 
everywhere, setting a limit on the density of 
diffuse, $z$-filling gas.}

\end{figure}

\end{document}